\documentclass{article}
\usepackage[utf8]{inputenc}
\pdfoutput=1  % togliere il commento se le figure sono pdf, Blasciarlo se sono eps
\usepackage{jcappub}
\usepackage{graphicx}
\usepackage{amsmath,amssymb}
\usepackage{url}
\usepackage{color} 
\usepackage{enumitem}
\usepackage{subfigure}

\newcommand{\exclude}[1]{}

\newcommand{\be}{\begin{equation}}
\newcommand{\ee}{\end{equation}}
\newcommand{\beq}{\begin{equation}}
\newcommand{\eeq}{\end{equation}}
\newcommand{\bea}{\begin{eqnarray}}
\newcommand{\eea}{\end{eqnarray}}

\newcommand{\bx}{{\mathbf  x}}

\newcommand{\bk}{{\mathbf k}}

\newcommand{\dd}{\partial}

\newcommand{\al}{\alpha}

\newcommand{\de}{\delta}
\newcommand{\ep}{\epsilon}

\newcommand{\la}{\lambda}

\newcommand{\si}{\sigma}

\newcommand{\Om}{\Omega}

\definecolor{dgreen}{rgb}{0,0.6,0.0}

\title{Magnetogenesis from early structure formation due to Yukawa forces.}

\author{Ruth Durrer and}
\affiliation{Department of Theoretical Physics, Universit\'e de Gen\`eve, Quai
E. Ansermet 24 , Gen\`eve, 1211, Switzerland.}
\emailAdd{ruth.durrer@unige.ch}	
\author{Alexander Kusenko} 
\affiliation{Department of Physics and Astronomy, University of California, Los Angeles \\ Los Angeles, California, 90095-1547, USA}
\affiliation{Kavli Institute for the Physics and Mathematics of the Universe (WPI), UTIAS \\The University of Tokyo, Kashiwa, Chiba 277-8583, Japan} 
\affiliation{Theoretical Physics Department, CERN, 1211 Geneva 23, Switzerland}
\emailAdd{kusenko@g.ucla.edu}	
\abstract{
Yukawa interactions can mediate relatively long-range attractive forces between fermions in the early universe.  
Such a globally attractive interaction creates an instability that can result in the growth of structure in the affected species even during the radiation dominated era.  The formation and collapse of fermionic  microhalos can create  hot fireballs at the sites of the collapsing halos which inject energy into the cosmic plasma.  In this paper we study a new phenomena which can take place in such models. We show that the injected energy can be partially converted into primordial magnetic fields and we estimate the correlation scale and the power spectrum of these fields. We show that they  may be the seeds of the observed astrophysical magnetic fields. 
}
\begin{document}

\maketitle
	
%%%%%%%%%%%%%%%%%%%%%%%%%%%%%%%%%%%

\section{Introduction}

The origin of astrophysical magnetic fields is not well understood.  The observed fields in galaxies and clusters can be explained by the dynamo amplification~\cite{Ruzmaikin} of  seed fields produced in the early universe or at later time during structure formation~\cite{Durrer:2013pga}.
However, there are convincing arguments that small magnetic fields of the order of $10^{-16}$~Gauss are also present in voids~\cite{Neronov:2010gir,Essey:2010nd}. Such fields are very difficult to produce at late times and are therefore most certainly primordial. Future observations will probe the structure of the magnetic fields in the intergalactic voids, where they are expected to remain closely related  to the primordial seed fields.  Combined with a theoretical understanding of magnetogenesis, this may open a new window on the early universe. 

In this paper we discuss a new class of scenarios in which magnetic fields with astrophysically relevant correlation lengths are generated.  It is based on a recently discovered phenomenon of primordial structure formation in some specific species~\cite{Amendola:2017xhl,Savastano:2019zpr, Casas:2016duf,Flores:2020drq, Domenech:2021uyx, Flores:2021jas,Flores:2022uzt,Domenech:2023afs}.  
Such early structure formation can be caused by some relatively long-range forces, for example, the Yukawa forces~\cite{Flores:2020drq, Domenech:2021uyx, Flores:2021jas,Flores:2022uzt,Domenech:2023afs}, which exhibit an instability similar to the gravitational instability, only stronger.  
The growth of structure in some species already during the radiation dominated era represents a possible new epoch in the history of the universe, and it can have profound implications for production of primordial black holes~\cite{Flores:2020drq,Flores:2021jas}, generation of the matter-antimatter asymmetry~\cite{Flores:2022oef}, the production of dark matter in the form of weakly interacting particles~\cite{Flores:2023nto}, etc.  In the present paper we show that in these scenarios also the generation of primordial magnetic fields
can take place during the epoch of early structure formation.

We illustrate the new paradigm with an example that uses a dark sector consisting of two particles: a fermion $\psi$ and a boson $\phi$. As is commonly assumed in models of asymmetric dark matter~\cite{Petraki:2013wwa,Zurek:2013wia}, the fermions $\psi$ are expected to develop an asymmetry similar to the baryon asymmetry of the universe via some high-scale interactions connecting the two sectors.  
We introduce a Yukawa interaction between $\psi$ and $\phi$.  On  length scales smaller than the mass of $\phi$, this Yukawa interaction represents a long-range attractive force which causes the formation of fermionic halos even  in a radiation dominated universe~\cite{Amendola:2017xhl,Savastano:2019zpr, Casas:2016duf,Flores:2020drq, Domenech:2021uyx, Flores:2021jas,Flores:2022uzt,Domenech:2023afs}. The attractive, relatively long-range Yukawa forces create an instability, which is similar to the gravitational instability, only much stronger. The growth and collapse of the halos of fermions can lead to bound states~\cite{Wise:2014jva,Gresham:2017cvl,Gresham:2018rqo}, or, thanks to  radiative cooling by the same Yukawa interactions, it can even lead to formation of primordial black holes~\cite{Flores:2020drq,Flores:2021jas}. We shall call this possibility {\it scenario~1}. Alternatively, the halos can disappear again due to fermion annihilation in the dense halos. 
 We shall call this possibility {\it scenario~2}.
 
If the particles $\phi$ and $\psi$  interact with standard model particles, the formation, collapse, and decay of the halos can locally inject energy into the ambient plasma. 
In this paper we  show that such an inhomogeneous energy injection into the cosmic plasma creates suitable conditions for magnetogenesis~\cite{Durrer:2013pga}. 

Let us illustrate the scenario with a simple model that involves one fermion and one boson interacting via a Yukawa coupling:
\begin{equation}
\mathcal{L} \supset \frac{1}{2}m_{\phi}^2{\phi}^2 - y{\phi}\bar{\psi}\psi + \cdots
.
\end{equation}

As discussed in Refs.~\cite{Amendola:2017xhl,Savastano:2019zpr, Casas:2016duf,Flores:2020drq, Domenech:2021uyx, Flores:2021jas,Flores:2022oef}, the growth of structures in $\psi$ particles due to the Yukawa instability, once it begins, proceeds very rapidly and reaches the nonlinear regime within a Hubble time $H^{-1}$. 
 Each halo collapses and virializes, but scalar bremsstrahlung ($\psi \psi \rightarrow \psi \psi \phi$) provides radiative cooling and removes  energy from the halo, allowing it to collapse further.  The outcome of this process depends on the presence of fermion number asymmetry in the dark sector.  In the absence of an asymmetry ($n_\psi=n_{\bar \psi}$), annihilations $\bar \psi \psi \rightarrow \phi \phi$ proceed at a growing rate and eventually decimate the halo ({\it scenario~2}).  Formation and disappearance of halos in this case can set the right conditions for baryogenesis~\cite{Flores:2022oef}.  Alternatively, if the dark fermions develop an asymmetry, which can be co-generated with the baryon asymmetry (as in the models of asymmetric dark matter~\cite{Petraki:2013wwa,Zurek:2013wia}), the outcome of halo collapse is the formation of primordial black holes (PBH), which can be dark matter ({\it scenario~1}).  The dark matter abundance is natural in the case of GeV scale particles with an asymmetry comparable to the baryon asymmetry of the universe. More precisely, using $\Om_b\simeq 0.2\Om_{\rm DM}$,
 
 \begin{equation}
\Omega_{\rm PBH} \simeq
0.2\, {\Omega_{\rm DM}}  \frac{m_\psi}{m_p} 
\frac{\eta_\psi}{\eta_{\rm B}}= {\Omega_{\rm DM}}   \left ( \frac{m_\psi}{5 \, {\rm GeV}} \right ) \left ( \frac{\eta_\psi}{10^{-10}} 
\right ) .
\label{eq:fpbh}
\end{equation}

Even in the case of co-generation of the asymmetry in the baryonic and the dark sector, the two parameters, $\eta_B$ and  $\eta_\psi$ may differ by a few orders of magnitude due to the differences in the temperatures and the numbers of degrees of freedom in the visible and the dark sectors, respectively~\cite{Petraki:2013wwa,Zurek:2013wia}.  The formation of structures in $\psi$ particles is possible after the fermions kinetically decouple from the plasma, which happens at temperature $T_*\sim 10^{-2} m_\psi$~\cite{Graesser:2011wi,Flores:2020drq}.   In what follows, we will assume the following representative values of the parameters, which are consistent with dark matter in the form of PBHs: 
\begin{eqnarray}
T_* &\sim & 1-100\ {\rm MeV}, \\
m_\psi & \sim & 0.1-10 \ {\rm GeV}.  
\end{eqnarray}

\section{Generation of turbulence}

It is reasonable to assume that the dark sector $\{\psi,\phi \}$ is weakly coupled to the standard model via some higher-dimensional operators or small kinetic mixing, which is, in fact, necessary for the co-generation of the fermion asymmetry~\cite{Petraki:2013wwa,Zurek:2013wia}. This implies that during the halo collapse and emission of scalar particles, a small fraction of the halo energy is transmitted to the cosmic plasma. In {\it scenario~1}, where the halos collapse to form the dark matter, the energy transferred to the plasma is a fraction of the dark matter density, $\ep_1\rho_{DM}$, while in {\it scenario~2} it is a faction of the radiation density, $\ep_2\rho_r$. 
This energy injection  results in the creation of a fireball at the site of each halo.  As the fireballs expand, since the Reynolds number of the plasma and the conductivity are very high, a significant fraction of this inhomogeneously deposited energy will be converted into MHD turbulence. We denote this fraction $f$. For completeness, we also present estimates of the relevant Reynolds numbers in Appendix~\ref{a:Reynold}.
We now parameterize the energy in turbulence as a fraction of the dark matter energy:
$\rho_{\rm K} = \epsilon_1f \, \rho_{\rm DM}=\ep \rho_{\rm DM}$ in {\it scenario 1}, while $\rho_{\rm K} = \epsilon_2f \, \rho_r=\ep \rho_r$ in {\it scenario 2}.
This is the only input of the dark sector physics really needed here to estimate the induced magnetic field, its amplitude and its spectrum: Early structure formation leads to turbulence in the cosmic plasma at some temperature $T_*$.

As electrons and positrons are still relativistic at $T_*$, when turbulence is generated, we can assume the turbulence to develop within a Hubble time into Kolmogorov turbulence with a correlation scale of the order of the Hubble scale.

During the radiation era,
\bea
H(T_*) &=& \frac{1}{2\tau_*} ~\simeq~ 0.2\sqrt{g_{\rm eff\, *}}\left(\frac{T_*}{1{\rm MeV}}\right)^2 {\rm sec}^{-1}\\
\la_* &=& \eta\frac{1+z_*}{H(T_*)} ~\simeq~ \eta\frac{1.2}{ \sqrt{g_{\rm eff\, *}}}10^{11}\left(\frac{1{\rm MeV}}{T_*}\right){\rm sec} ~\simeq ~ 
\eta\frac{1.3}{ \sqrt{g_{\rm eff\, *}}}\left(\frac{1{\rm MeV}}{T_*}\right){\rm kpc} \,.
\eea
Here $\tau$ is cosmic time and $\la_*$ is the {\em comoving} correlation scale at the time $\tau_*$, at temperature $T_*$ and  $g_{\rm eff\,*}$ is the number of relativistic degrees of freedom at $T_*$. The speed of light is set to $c=1$ and we normalize the scale factor to one today, $a_0=1$. The factor $\eta\lesssim 1$ is ratio $\la_*/[(1+z_*)H_*^{-1}]$.
We assume that rapidly (i.e. within less than one Hubble time), on wave numbers $k$ with  $2\pi/\la_*=k_* \leq k\leq k_d$ a Kolmogorov spectrum is established.
The scale $k_d$ is a damping scale beyond which the turbulent motion is dissipated into heat.

 Furthermore, the relativistic plasma is compressible, $\dd P/\dd\rho = 1/3$, hence compressible turbulence with a white noise spectrum on large scales, $k<k_*=2\pi/\la_*$  develops. The turbulent velocity
power spectrum of causally generated compressible turbulence can be approximated as follows
\bea
\langle v_i(\bk)v_j(\bk')\rangle  &=& (2\pi)^3\de(\bk-\bk')\de_{ij}P_K(k)
\eea
with
\bea
P_K(k) &=&   v_*^2k_*^{-3}\times\left\{ \begin{array}{ll}1\,, & k<k_* \\
 (k/k_*)^{-11/3} \,, & k_*\leq k < k_d \qquad\\
 0 \,, &  k_d \leq k \,.
\end{array}\right.
\eea

Even though Kolmogorov has derived the spectral index $-11/3$ only for non-relativistic turbulence, it has been shown in numerical simulations that also relativistic turbulence develops a (nearly) Kolmogorov spectrum~\cite{Zhdankin:2016lta,Comisso:2018kuh}.
The amplitude $v_*^2$ is proportional to  $\ep$.

The energy density in the turbulence is given by the integral of the velocity power spectrum,
\bea
\rho_K &\simeq& \frac{\rho_r }{2}\langle v(x)^2\rangle ~
= ~  \frac{3\rho_r}{4\pi^2}\int P_K(k)k^2 dk ~ = ~  \frac{11}{8\pi^2}\rho_rv_*^2 \,.
\eea
Here $\rho_r$ is the energy in relativistic particles which dominates the total energy density in the radiation era.

\section{Generation of  magnetic fields}
 As the cosmic plasma is charged and has very high conductivity, the turbulence becomes MHD turbulence and will also lead to the generation of magnetic fields~\cite{LL8,Biskamp}.
Equi-partition dictates that the magnetic field spectrum is of the form
\be
\langle B_i(\bk)B_j^*(\bk')\rangle = (2\pi)^3\de(\bk-\bk')\left(\de_{ij}-\hat k_i\hat k_j\right)P_B(k) \label{ePB}
\ee
with
\be
P_B(k) =   B_*^2k_*^{-3}\times\left\{ \begin{array}{ll}(k/k_*)^2 \,, & k<k_* \\
(k/k_*)^{-11/3} \,, & k_*\leq k < k_d \\
 0 \,, &  k_d \leq k \,.
\end{array}\right.
\ee
The $k^2$ behavior on large scales is due to the fact that the spectrum must be analytic for small $k$ as the correlation function in real space has compact support~\cite{Durrer:2003ja} (there are no correlations on super-horizon scales). On intermediate scales, $k_*\leq k < k_d$ the spectral index is the one of Kolmogorov turbulence while on small scales, $k>k_d$ the field is dissipated into heat.
Note that we always calculate the {\em comoving} magnetic field. The true magnetic field is $B/a^2$.
The energy density in this magnetic field is given by $\rho_B$, where (we use Heavyside units so that $B$ is canonically normalized)
\bea
a^4\rho_B(\tau_*) &=& \frac{1}{2}\langle B(\bx)^2\rangle =  \frac{1}{2\pi^2}\int_0^\infty P_B(k)k^2dk  \nonumber \\
&=& \frac{1/5 + 3/2}{2\pi^2}B_*^2 = \frac{17 }{20\pi^2}B_*^2\,.
\eea
Typically we expect equi-partition, $\rho_B\simeq \rho_K$. 
 
We want to express $\ep$ in terms of the magnetic field $B_*$ for both scenarios.  The ratio of the energy density in magnetic fields and in thermal radiation of $g_{\rm eff}$ degrees of freedom at temperature $T_*$ is given by
\be
\frac{\rho_B}{\rho_r} = \frac{a^4\rho_B}{a^4\rho_r} \simeq \frac{\langle B^2\rangle }{g_{\rm eff}\left(2\times 10^{-6} {\rm Gauss}\right)^2} \,.
\ee
Therefore, for {\it scenario 2}
\be
\ep \simeq  \frac{\rho_B}{\rho_r} \simeq 10^{-18}\left(\frac{B_*}{10^{-15}{\rm Gauss}}\right)^2\,.
\ee
For {\it scenario 1} we need the ratio of $\rho_B$ and $\rho_{DM}$. 
At the time $\tau_*$, the Universe is radiation dominated, hence the fraction of the dark matter density  dumped into magnetic fields can be expressed as
\bea
\hspace*{-0.2cm}\frac{\rho_B(\tau_*}{\rho_{\rm DM}(\tau_*)} &\simeq& \frac{\rho_B}{\rho_r}\frac{1+z_*}{1+z_{eq}} \simeq \frac{\rho_B}{\rho_r}\frac{T_*/T_0}{3200}
\nonumber \\  
&\simeq& 1.3\times 10^6\left(\frac{T_*}{1{\rm MeV}}\right)\frac{\langle B^2\rangle }{g_{\rm eff}\left(2\times 10^{-6} {\rm Gauss}\right)^2} 
%\nonumber \\
%&\simeq&  
~\simeq ~ 10^{-12}\left(\frac{T_*}{1{\rm MeV}}\right)\left(\frac{B_*}{10^{-15}{\rm Gauss}}\right)^2,
\eea
were $T_0=2.73$K is the present temperature of the Universe and $1+z_{eq} \simeq 3200$ is the redshift of equal matter and radiation.
Hence for  $B_* =10^{-15}$ Gauss one only needs $\ep\sim 10^{-18}$ in {\it scenario~2}. In {\it scenario~1}, the value of $\ep$ also depends on the temperature. At $T_*\sim 1$MeV one needs $\ep\sim 10^{-12}$, while the same amplitude magnetic fields at $T_*\sim 10^5$GeV requires $\ep\sim 10^{-4}$.

If the magnetic fields would simply scale like $a^{-2}$, $B_*$ would be the physical magnetic field today. However, MHD turbulence once generated is freely decaying with an  inverse cascade due to the white noise velocity power spectrum.
 This leads to significant changes in both, the amplitude of the magnetic field and its correlation scale.which can grow up to nearly 1Mpc.
 
At $T_*$ the situation is as discussed above. Let us now study its evolution.
Equi-partition requests that $v_*\simeq v_A = \sqrt{\rho_B/\rho_r}$, where $v_A$ is the Alfvén velocity.

\begin{figure}[!ht]
\begin{center}
\includegraphics[width=9cm]{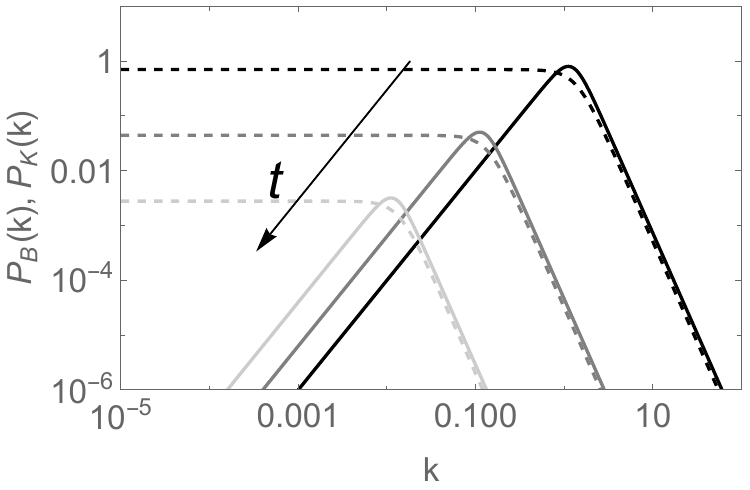}
\end{center}
\caption{The evolution of the magnetic field (solid lines) and the velocity (dashed lines) spectra in an MHD plasma with compressible turbulence in arbitrary units. Lighter grey scales indicate later times. The spectra evolve towards smaller $k$. The decay of the peak amplitude is due to dissipation which is discussed below (see Eq.~\eqref{e:Bevol}).}
\label{f:Bevol}
\end{figure}

While some mechanisms can generate helicity of primordial magnetic fields, and it may be detectable~\cite{Vachaspati:1991nm,Cornwall:1997ms,Kahniashvili:2005yp,Durrer:2010mq,Chen:2014rsa,Chen:2014qva,Long:2015bda}, 
we expect the magnetic fields generated in our scenario to be non-helical as the turbulent energy injected in the cosmic plasma has no preferred helicity, in the simplest models which we consider here. Nevertheless, since the relativistic plasma is compressible, a mild inverse cascade is generated.
The comoving power spectra  and the correlations scale $k_*$ evolve with time (see \cite{Durrer:2013pga}  where this is derived in detail). More precisely, $\la_*$ behaves as
\be\label{e4:laBn}
\lambda_*(t)\propto t^{\frac{2}{5}}  \,.
\ee
Here we use the fact that we consider compressible turbulence, where the velocity has a white noise spectrum. Therefore, the correlation scale grows significantly as equi-partition requires the peak of the magnetic field power spectrum to move to the left (see Fig.~\ref{f:Bevol}), provoking an inverse cascade.   The energy density evolves as
\be
a^4 \rho_B\simeq a^4\rho_K\propto \frac{\overline v^2}{2} \propto  t^{-6/5}
\ee
Here and above, $t$ is {\em conformal} time. Also this result is derived in detail in~\cite{Durrer:2013pga}  and has also been confirmed by numerical simulations~\cite{Hindmarsh:2013xza}.
Schematically the evolution looks as shown in Fig.~\ref{f:Bevol}.

This evolution continues until about recombination  after which there is no longer a charged plasma and the evolution (roughly) stops.

 At late times we find a correlation scale of
\bea
\la_*(t_0) &\simeq& \la_*(t_{\rm rec}) = \la_*(t_*)\left(\frac{t_{\rm rec}}{t_*}\right)^{2/5} 
~ \simeq ~  \la_*(t_*)\left(\frac{z_*+1}{z_{\rm rec}+1}\right)^{2/5} ~ \simeq   ~   \la_*(t_*)\left(\frac{(T_*/1{\rm MeV})10^{10}}{2.35\times1000}\right)^{2/5} \nonumber \\
 &\simeq&  450\times \la_*(t_*)\left(\frac{T_*}{1{\rm MeV}}\right)^{2/5}  ~ \simeq ~   0.6{\rm Mpc}\frac{\eta}{\sqrt{g_{\rm eff}}}\left(\frac{1{\rm MeV}}{T_*}\right)^{3/5} \,.
\eea
Correspondingly one finds for the evolved magnetic field
\bea\label{e:Bevol}
B_0 &\equiv& B_*(t_0) \simeq	 B_*(t_{\rm rec}) \simeq B_*(t_*) \left(\frac{t_{\rm rec}}{t_*}\right)^{-3/5}  
~ \simeq ~  B_*(t_*)\left(\frac{z_*+1}{z_{\rm dec}+1}\right)^{-3/5} \\ 
&\simeq&  B_*(t_*)\left(\frac{(T_*/1{\rm MeV})10^{10}}{2.35\times1000}\right)^{-3/5} %\nonumber \\
 ~ \simeq~    10^{-4}\times B_*(t_*)\left(\frac{T_*}{1{\rm MeV}}\right)^{-3/5}\,.
\eea
Hence, for $T_*\sim 1$MeV, in order to have magnetic fields of about $10^{-16}$Gauss today, we need $B_*(t_*)\simeq 10^{-12}$ Gauss. In general, $\ep$ relates to the {\em present} magnetic field strength at the correlation scale, $B_0$, as follows:
\bea
 && \mbox{\it Scenario 1 :}\nonumber \\ 
\ep=\frac{\rho_B(\tau_*)}{\rho_{\rm DM}(\tau_*)} &\simeq& 10^{-6} \left(\frac{T_*}{1{\rm MeV}}\right)^{11/5}\left(\frac{B_0}{10^{-16}{\rm Gauss}}\right)^2\,, \\
 && \mbox{  }\nonumber \\ 
 && \mbox{\it Scenario 2 :}  \nonumber \\ 
\ep =\frac{\rho_B(\tau_*)}{\rho_{r}(\tau_*)} &\simeq& 10^{-12} \left(\frac{T_*}{1{\rm MeV}}\right)^{6/5}\left(\frac{B_0}{10^{-16}{\rm Gauss}}\right)^2 \,.
\eea
Here $B_0$ is the (physical) magnetic field amplitude today.

\begin{figure}[!ht]
\begin{minipage}{6cm}
\includegraphics[width=6cm]{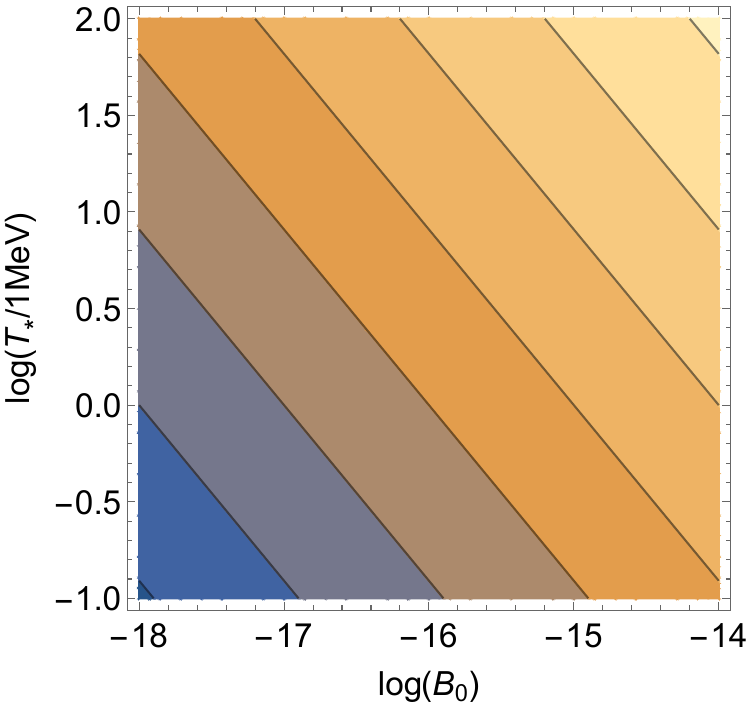}
\end{minipage}
\begin{minipage}{1cm}
\includegraphics[width=1cm]{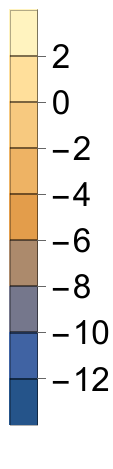}
\vspace*{1cm}
\end{minipage}
\hspace*{0.2cm}
\begin{minipage}{6cm}
\includegraphics[width=6cm]{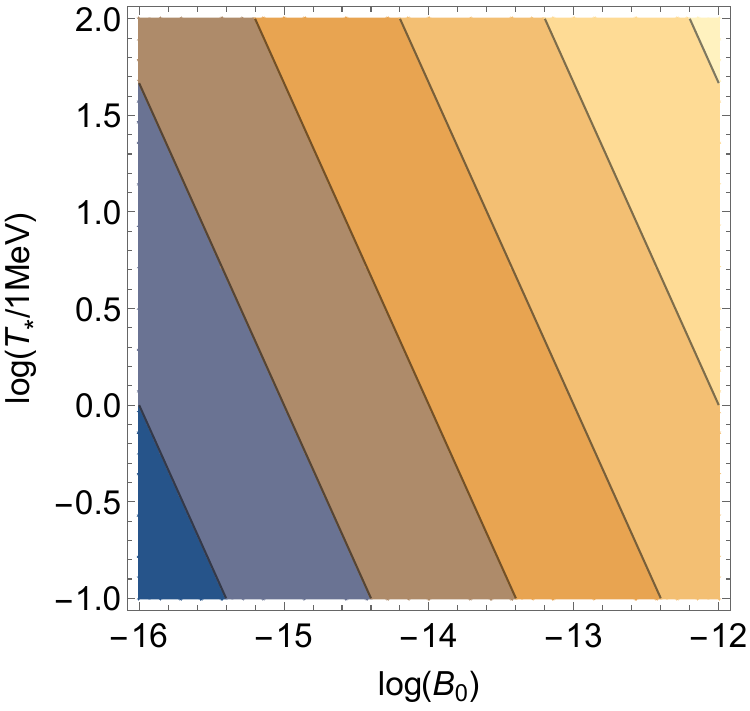}
\end{minipage}
\begin{minipage}{1cm}
\includegraphics[width=1cm]{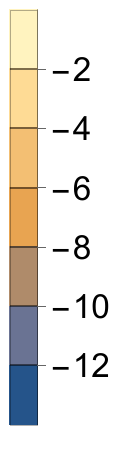}
\vspace*{1cm}
\end{minipage}
\caption{The left panel shows the required ratio, $\log\ep =\log\left(\frac{\rho_B(\tau_*)}{\rho_{\rm DM}(\tau_*)}\right)$  as a function of temperature $T_*$ and final magnetic field $B_0$ (the log's are to base 10) for  {\it scenario~1}. \newline
The right panel shows the required ratio, $\log\ep =\log\left(\frac{\rho_B(\tau_*)}{\rho_{r}(\tau_*)}\right)$ as a function of temperature $T_*$ and final magnetic field $B_0$ (the log's are to base 10) for  {\it scenario~2}.
\label{f:rhorat}}
\end{figure}

  In Fig.~\ref{f:rhorat} we plot the required initial ratio of $\ep=\rho_B/\rho_{\rm DM}$ for {\it scenario~1} (top panel) and $\ep=\rho_B/\rho_{r}$  for {\it scenario~2} (bottom panel) needed to obtain a field amplitude $B_0$ today for a given explosion temperature $T_*$. For  {\it scenario~1}, the two brightest colors with $\ep\geq 1$ are clearly excluded. We must require at least $\ep<0.1$. A present amplitude of about $10^{-14}$Gauss with a formation temperature of $1$MeV would require $\ep=0.01$ which seems reasonable. Also an amplitude of $10^{-16}$ to  $10^{-15}$ Gauss at $T_*=10$MeV to $T_*=1$MeV can be obtained with a modest value of $\ep\simeq 10^{-4}$. 
The results for  {\it scenario~2} are much more optimistic. For all the injection temperatures considered here, $0.1$MeV$\leq T_*\leq 100$MeV present magnetic fields of nearly up to $10^{-12}$Gauss can be  generated with $\ep<10^{-2}$. For a present magnetic field of $10^{-16}$Gauss even a value of $\ep\leq 10^{-9}$ is sufficient.

Considering Fig.~\ref{f:Bevol} it is clear, that the spectrum of the expected magnetic fields is very similar to the one of a first order phase transition. It is determined by causality on  scales larger than the correlation length and by Kolmogorov turbulence and dissipation at smaller scales. These plasma processes are independent of the mechanism that injects the energy into the plasma. The question which poses itself is whether this scenario can be distinguished observationally from a first order phase transition which also can inject energy into the cosmic plasma leading to turbulence and magnetogenesis along the same lines. The most prominent examples are the electroweak (EW) and the confinement (QCD) phase transitions. The first one is at much higher temperature, $T_{\rm EW}\simeq 200$GeV and would therefore lead to a much smaller correlation length. Furthermore, in most models which lead to a first order EW phase transition, this transition generates helical magnetic fields. The situation is more difficult with the QCD transition. If this transition is first order (this might be achieved e.g. with a high chemical potential in the lepton sector~\cite{Schwarz:2009ii}), it happens only somewhat above the upper limit of the temperatures considered here, $T_{\rm QCD}\sim 150$MeV, which would be difficult to distinguish from  models with  $T_*\sim 100$MeV.
However, if $T_*\sim 1$ to $10$MeV, there is no cosmological phase transition expected in this regime and our model might be more natural.
Furthermore, the telltale signature of the first scenario would of course be magnetic fields in combination with primordial black holes in the expected mass range. As always in cosmology, we want to collect several observable 'relics'  from a previous event before we take a model seriously.

\section{Conclusions}
In this work we have found that explosions such as they are expected from early structure formation in the radiation dominated Universe at $100$MeV $>T_*> 0.1$MeV can not only lead to primordial black holes, but also to magnetic fields with an amplitude of about $10^{-16}$Gauss or more. The correlation scale of these fields can be of order Mpc for $T_*= 1$MeV and somewhat smaller/larger at higher/lower temperatures. A significantly lower temperature is most probably in conflict with other observations such as primordial nucleosynthesis. In the first scenario, where primordial black holes are formed which are the dark matter, explosions at a much higher temperature than the above interval generate magnetic fields that have too low amplitude today due to dilution and turbulent decay.
In the second scenario where the dark sector fermions annihilate into radiation, magnetic fields with amplitude up to $10^{-16}$Gauss can be generated even at $T_*\sim 10^6$GeV, albeit with much shorter correlation length. Note also, that the important ingredient here was just that we have fast energy injection into the cosmic plasma at some high temperature $T_*$, correlated over a scale close to the Hubble scale $H^{-1}(T_*)$,
and that we can characterize the injected energy as $\ep\rho_{\rm DM}$ ({\it scenario~1}) or $\ep\rho_{r}$ ({\it scenario~2}). The details of the energy injection mechanism are irrelevant for these results.

If, indeed, the seed magnetic fields are produced in the early universe and are amplified in galaxies and clusters, the fields deep in the voids are expected to retain information about the original primordial magnetic power spectrum. The upcoming gamma-ray observations with the Cherenkov Telescope Array (CTA) can provide new information because the gamma-ray cascades from blazars are sensitive to both the amplitude and the correlation lengths of the magnetic  fields in the intergalactic voids~\cite{Neronov:2006lki,Neronov:2010gir,Essey:2010nd,Dermer:2010mm,Razzaque:2011jc}.  Thus future observations may open a new window on the early universe and elucidate the processes that generated primordial magnetic seed fields. 
\vspace{0.2cm}

\begin{acknowledgments}
We thank Guillem Dom\`enech, Marcos Flores, Derek Inman, Andrii Neronov, and Misao Sasaki for helpful discussions.
R.D. acknowledges support from the Swiss National Science Foundation, grant No. 200020\underline{~}182044. 
A.K. was supported by the U.S. Department of Energy (DOE) Grant No. DE-SC0009937, by the Simons Foundation Fellowship, by the World Premier International Research Center Initiative (WPI), MEXT, Japan, by Japan Society for the Promotion of Science (JSPS) KAKENHI grant No. JP20H05853, and by the UC Southern California Hub with funding from the UC National Laboratories division of the University of California Office of the President.
\vspace{2cm}
\end{acknowledgments}

%\newpage

\appendix

{\LARGE\bf APPENDIX}
\vspace{1cm}

\section{Reynolds numbers in the cosmic plasma}\label{a:Reynold}
In this section we present estimates of the Reynolds numbers in the cosmic plasma at temperatures $1$MeV $\leq T\leq 100$MeV for completeness. The results of this appendix are not new but also not widely known. More details are found in Refs.~\cite{Caprini:2009pr,Caprini:2009yp,Durrer:2013pga}. We mainly follow the presentation in~\cite{Caprini:2009yp}.

The kinetic Reynolds number of a plasma at a scale $\la$ is given by
\be
{\rm Re_{kin}}(\la,T) = \frac{v_\la \la}{\nu(T)}
\ee
where $v_\la$ is the typical velocity at scale $\la$ and $\nu(T)$ is the kinematic viscosity. At $100$MeV$>T>1$MeV, viscosity is dominated by the neutrinos which are the most weakly coupled particles still in equilibrium at these temperatures leading to
\be
\nu =\frac{\ell_{\rm mfp}}{5} \simeq \frac{1}{15G_F^2 T^5}\,,
\ee
where $\ell_{\rm mfp}$ denotes the mean free path and $G_F=(293$GeV$)^{-2}$ is the Fermi constant.
Inserting numbers one can express $\nu$ as
\be
\frac{1}{\nu} \simeq 3\times 10^6H(T)\left(\frac{T}{100{\rm MeV}}\right)^3, \qquad 1{\rm MeV}< T< 100{\rm MeV}\,,
\ee
where $H(T)$ is the Hubble scale at temperature $T$. For fireballs of size a roughly the Hubble scale, $\la \sim  H^{-1}$ at relativistic speed, $v_\la \sim 1$, we have a  high Reynolds number hence turbulence develops. Note that our approximation becomes inaccurate close to the lower bound, $T\sim 1$MeV since neutrinos decouple at roughly $1.4$MeV and for $T\lesssim 3$MeV also the viscosity of photons has to be considered which is much smaller. Therefore, even at $T\sim 1$MeV the kinetic Reynolds number remains significantly larger than 1 for relativistic motions at the Hubble scale, see~\cite{Durrer:2013pga} for more details.

The magnetic Reynolds number is given by the same expression as the kinetic one, replacing kinetic viscosity by magnetic diffusivity which is the inverse of the conductivity, $\si$. To confirm that also the magnetic Reynolds number ${\rm Re_m}$ is much larger than one, it suffices to determine the Prandl number, $P_m$, which is defined as 
\be
P_m(T) = \frac{{\rm Re_m}}{{\rm Re_{kin}}} =\si(T)\nu(T)\,,
\ee
and to show that $P_m>1$ in the temperature range of interest.
Using the expression for the conductivity of a relativistic electron-positron plasma derived in~\cite{Caprini:2009yp}
\be
\si(T) \simeq  \frac{T}{\al(T)}\,,
\ee
where $\al(T)$ is the fine structure constant at the temperature $T$, we obtain
\be
P_m(T) \simeq \frac{137}{15G_F^2T^4} \sim 5\times 10^{14}\left(\frac{100{\rm MeV}}{T}\right)^4\gg 1\,, \qquad 1{\rm MeV}< T< 100{\rm MeV}\,.
\ee
%%%%%%%%%%%%%%%%%%%%%%%%%%%%%%%%%%%% 
 \bibliographystyle{JHEP}
\bibliography{biblio}

\end{document}